\documentclass[twocolumn,showpacs,preprintnumbers,superscriptaddress,amsmath,amssymb,prl]{revtex4}

\usepackage{graphicx,epsfig,color} 
\usepackage{dcolumn}
\usepackage{bm}
\usepackage{amsmath}
\usepackage{indentfirst}
\usepackage{url}

\newcommand{\be}{\begin{equation}}
\newcommand{\ee}{\end{equation}}
\newcommand{\ba}{\begin{eqnarray}}
\newcommand{\ea}{\end{eqnarray}}

\begin{document}

\preprint{APS preprint}

\title{The 2006-2008 Oil Bubble and Beyond}
 
\author{D. Sornette}
\affiliation{Chair of Entrepreneurial Risks, Department of
Management, Technology and Economics, ETH Zurich, CH-8001 Zurich,
Switzerland \\}
\affiliation{Swiss Finance Institute, c/o University of Geneva, 40 blvd. 
Du Pont dÕArve, CH 1211 Geneva 4, Switzerland}

\author{R. Woodard}
\affiliation{Chair of Entrepreneurial Risks, Department of
Management, Technology and Economics, ETH Zurich, CH-8001 Zurich,
Switzerland \\}

\author{W.-X. Zhou}
\affiliation{School of Business, School of Science, Research Center
for Econophysics and Research Center of Systems Engineering, East
China University of Science and Technology, Shanghai 200237, China}

\date{\today}

\begin{abstract}

  We present an analysis of oil prices in US\$ and in other major
  currencies that diagnoses unsustainable faster-than-exponential
  behavior.  This supports the hypothesis that the recent oil price run-up
  has been amplified by speculative behavior of the type found during
  a bubble-like expansion. We also attempt to unravel the information
  hidden in the oil supply-demand data reported by two leading
  agencies, the US Energy Information Administration (EIA) and the International
  Energy Agency (IEA). We suggest that the found increasing
 discrepancy between the EIA and IEA figures provides a measure of
the estimation errors. Rather than
a clear transition to a supply restricted regime, we interpret the discrepancy between the 
IEA and EIA as a signature of uncertainty, and 
there is no better fuel than uncertainty to promote speculation!

\end{abstract}



\maketitle

Since 1995, the US markets have lived through three major episodes,
now recognized by most professionals and regulators and a growing
number of academics as bubbles: the new economy ICT
(Internet-Communication-Technology) frenzy culminating in 2000, the
real-estate surge peaking in the US in mid-2006 and the subprime NIV
(new instrument vehicle) boom, which topped in 2007. In finance and
economics, the term ÒbubbleÓ refers to a situation in which excessive
expectations of future price increases cause prices to be temporarily
elevated without justification from fundamental valuation.

Since approximately March 2008, a growing number of journalists,
pundits \cite{Soros}, bankers \cite{Credit_Suisse} and academics
\cite{Wharton,krugman} have been discussing the pros and cons of the
hypothesis that commodities, and in particular oil, have entered a
bubble regime.  One key question is to explain the quadrupling of oil
prices since 2003. Some attribute it mainly to the pricing of the
growing demand (in particular from the emergent China and India
markets) imperfectly balanced by the increasingly apparent limits of
world oil production. Others are raising the specter of rising
speculation \cite{Soros}.
 
\begin{figure}[h]
\includegraphics[width=8cm]{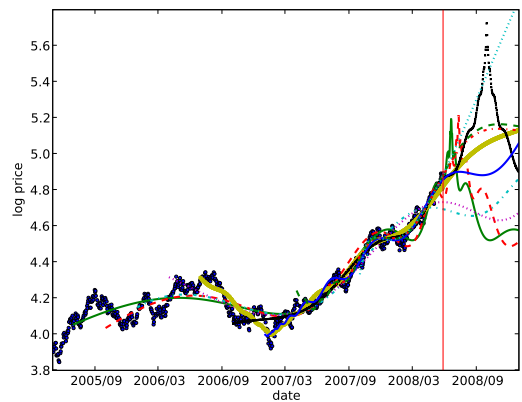}
\caption{\label{0_2805oilsubI_all_fits} Typical result of the
  calibration of the simple LPPL model to the oil price in US\$ in
  shrinking windows with starting dates $t_{\rm start}$ moving up
  towards the common last date $t_{\rm last}=$ May 27, 2008.  }
\end{figure}

Based on analogies with statistical physics and complexity theory, we
have developed in the last decade an approach that diagnoses bubbles
as transient super-exponential regimes \cite{Sorbookcrash}. In a
nutshell, our methodology aims at detecting the transient phases where
positive feedbacks operating on some markets or asset classes create
local unsustainable price run-ups. The mathematical signature of these
bubbles is a log-periodic power law (LPPL)
\cite{SJ1998,JS1999,JSL1999,JLS2000,SJ2001}.  The power law models the
faster-than-exponential growth culminating in finite time.  The
log-periodic oscillations reflect hierarchical structures
\cite{JSL1999,JLS2000} as well as competition between the trading
dynamics of fundamental value and momentum investors
\cite{Ide-Sornette}.

Here, we present a brief synopsis of an extended analysis that we have
performed to address the question of whether oil prices exhibit a
bubble-like dynamics, which may be symptomatic of speculative
behavior. We have obtained robust and reliable diagnostics (i) by
comparing different implementations of the LPPL theory, called the
simple LPPL model \cite{SJ2001}, the second-order Weierstrass model
\cite{Weirstrass} and the second-order Landau model
\cite{JS_Landau2,JS_Landau2-3,JS_japan_eval}, (ii) by performing
extensive sensitivity analyses with respect to many different time
windows used to calibrate the models and (iii) by using bootstrap
methods to resample the residues over monthly time scales so as to
keep as much as possible the statistical properties of the time series
in the bootstrap scenarios. In our detailed analysis, we condition the
calibration on a certain number of additional constraints that ensure
the statistical significance of the LPPL structure, which include
bounds on the key parameters informed from previous analyses
\cite{SJ2001,JS_endo-exo}, and the statistical significance of the
power law and log-periodic components \cite{ZS_stat_sign-02}.  In
addition, to address the question of a possible interplay between oil
price increase and US-dollar depreciation, we perform the same
analysis for oil price expressed in euro and in other major
currencies.

\begin{figure}[h]
  \includegraphics[width=8cm]{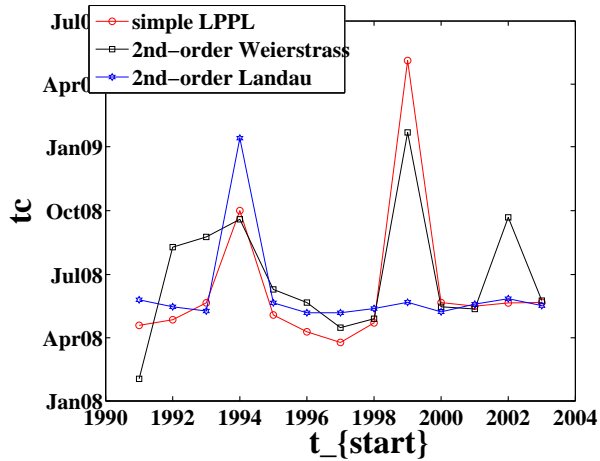}
  \caption{\label{Fig_2_tc} Predicted critical time $t_c$ obtained
    using the three LPPL models (simple LPPL, second-order Weierstrass
    and second-order Landau) as a function of the beginning time
    $t_{\rm start}$ for the fixed $t_{\rm last}=$ May 27, 2008.  }
\end{figure}

Figure \ref{0_2805oilsubI_all_fits} shows a typical result of the
calibration of the simple LPPL model to the oil price in US\$ in
shrinking windows with starting dates $t_{\rm start}$ moving up
towards the common last date $t_{\rm last}=$ May 27, 2008. One
particular useful feature of the LPPL models is that, in contrast with
most econometric models, they describe transient regimes ending at a
critical time $t_c$ beyond which the bubble is supposed to cross-over
to another regime, either by crashing or through a more progressive
transition \cite{JS_endo-exo,forecasting}.  Figure \ref{Fig_2_tc}
shows the predicted critical time $t_c$ obtained using the three LPPL
models (simple LPPL, second-order Weierstrass and second-order Landau)
as a function of the beginning time $t_{\rm start}$ for the fixed
$t_{\rm last}=$ May 27, 2008.  Extensive scanning of $t_{\rm start}$
and $t_{\rm last}$ confirms the main messages of figures
\ref{0_2805oilsubI_all_fits} and \ref{Fig_2_tc} of (a) a reliable
detection of a LPPL regime confirming the existence of a bubble in oil
price expressed in US\$ and (b) a robust and stable diagnostic that
the bubble is close to a local peak (and actually may have already
reached it). We cannot however exclude the possibility that the
proximity to a critical time $t_c$ is only a temporary process
embedded in a larger-scale bubble, that could develop in the coming
months and years.

\begin{figure}[h]
\includegraphics[width=8cm]{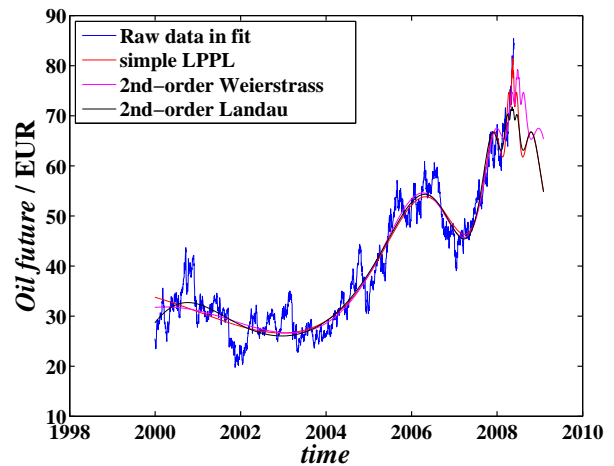}
\caption{\label{Fig_3_OILinEUR_10} Three fits with the simple LPPL,
  second-order Weierstrass and second-order Landau model of the oil
  price expressed in euro.  }
\end{figure}

Figure \ref{Fig_3_OILinEUR_10} shows the three fits with the simple
LPPL, second-order Weierstrass and second-order Landau model of the
oil price expressed in euro.  This confirms that the bubble is
genuine, and not solely a consequence of the weakening of the
US\$. The values of the critical time $t_c$ determined from these and
other calibrations in different time windows and using other major
currencies are found similar to those reported in figure
\ref{Fig_2_tc}, confirming the existence of a bubble phenomenon.  In
addition, our analysis points to a distinct change of regime in the
oil price dynamics in US\$ occurring between the last quarter of 2005
and the first quarter of 2006, beyond which a net acceleration can be
observed, perhaps correlated with the deregulation of Intercontinental
Exchange (ICE) oil futures in US markets by the U.S. Commodity Futures
Trading Commission.

\begin{figure}[h]
\includegraphics[width=8cm]{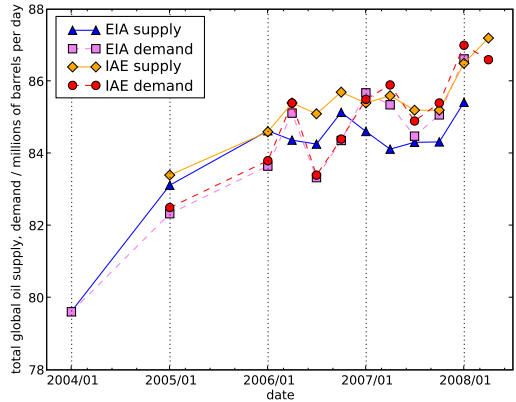}
\caption{\label{Fig_4_unnormalized_oil_price_supply_demand} Time series from 2004 to the first
quarter of 2008 of the World total liquid fuel demand and total World supply, 
as estimated by two agencies, the International Energy Agency (IEA) and the US Energy Information Administration (EIA) (\protect\url{http://www.eia.doe.gov/emeu/international/oilother.html}).}
\end{figure}

One last issue needs to be addressed: could the faster-than-exponential 
price rises demonstrated here result from a faster-than-exponential rise in demand which is not met by supply? If the answer is positive, our interpretation that we are seeing speculation unfolding
would be incorrect \cite{arXiv Blogger}. Could it indeed be that the recent price 
surges are explained for instance by a faster-than-exponential rise in demand from economies such as China and India? The recent paper \cite{JiangZemin} by former President Jiang Ze-Min himself
debunks this hypothesis at least for China (see Fig. 3 with caption in English in \cite{JiangZemin}).

To investigate this issue further, we took the values on World liquid fuel supply and demand reported by the International Energy Agency in its  May, 13, 2008 Oil Market report  \cite{IEA08} 
(see Table 1, p. 51) and by the US Energy Information Administration (EIA) (\url{http://www.eia.doe.gov/emeu/international/oilother.html}).
Figure \ref{Fig_4_unnormalized_oil_price_supply_demand} 
shows the World total liquid fuel demand and total World supply, 
as estimated by these two agencies (IEA and EIA).

While the two agencies report approximately consistent demand figures over this time period, there is a
more worrisome discrepancy between the supply values, with the EIA reporting a supply value about one Mb/d smaller than the IEA, since 2006. The EIA data suggests that
oil demand has exceeded supply over the last 5 quarters shown here,  suggesting that fundamentals 
play a major role in the price run up.  In contrast, the IEA data suggests a much weaker effect.
We tried to understand the causes of these different values. For one, each of these
estimated numbers aggregate global statistics coming from many sources and countries.
Second, there is also a degree of extrapolation and guess work, which is handled differently
in the two agencies. There are often revisions coming later (not unlike revision of GPD estimates
in macro-economics) that close the gap between these differences.  It seems to us
that one message is that the discrepancy between the EIA and IEA provides in fact a measure of
the estimation errors. In other words, these numbers are not to be believed at face value
given the uncertainties.

Given these 
uncertainties, one feature seems to emerge with a certain degree of certainty: until the end
of 2005, both agencies were in line and supply was systematically exceeding demand. Since 2006, this
deterministic fact has broken down with the ushering into an epoch of uncertainty. 
In our opinion, one should not conclude that demand has exceeded supply or vice-versa
since 2006, but rather that the oil market has entered an opaque regime. Rather than
a clear transition to a supply restricted regime, we interpret the discrepancy between the 
IEA and EIA as a signature of uncertainty. Here, we should
immediately stress that there is no better fuel than uncertainty to promote speculation!
 
In conclusion, the present study supports the hypothesis that the
recent oil price run-up, when expressed in any of the major
currencies, has been amplified by speculative behavior of the type
found during a bubble-like expansion.  The underlying positive
feedbacks, nucleated by rumors of rising scarcity, may result from one
or several of the following factors acting together: (1) protective
hedging against future oil price increases and a weakening dollar
whose anticipations amplify hedging in a positive self-reinforcing
loop; (2) search for a new high-return investment, following the
collapse of real-estate, the securitization disaster and poor yields
of equities, whose expectations endorsed by a growing pool of hedge,
pension and sovereign funds will transform it in a self-fulfilling
prophecy; (3) the recent development since 2006 of deregulated oil
future trading, allowing spot oil price to be actually more and more
determined by speculative future markets \cite{deregulation} and thus more
and more decoupled from genuine supply-demand equilibrium.

\vskip 0.5cm {\bf Acknowledgements}: We express our gratitude to
Didier Darcet from Renaissance Investment Management for useful discussions and Tam Hunt
from the Community Environmental Council of 
Santa Barbara for a stimulating correspondence.  WXZ acknowledges financial
supports from the National Natural Science Foundation of China (Grant
Nos. 70501011), the Fok Ying Tong Education Foundation (Grant
No. 101086), and the Program for New Century Excellent Talents in
University (Grant No. NCET-07-0288).


\begin{thebibliography}{}

\bibitem{Soros} Zumbrun, J., Soros tells congress to pop an oil
  bubble, Forbes, 3rd June 2008.

\bibitem{Credit_Suisse} Credit Suisse, The Investment Committee
  Meeting of 27 May 2008.

\bibitem{Wharton} Siegel, J. and W. Henisz, What's Behind the
  Flare-ups in Oil Prices? Jeremy Siegel and Witold Henisz Weigh In,
  Knowledge@Wharton, May 28, 2008.

\bibitem{krugman} Krugman, P., More on oil and speculation (The
  Conscience of a Liberal), The New York Times, May 13, 2008.

\bibitem{Sorbookcrash} Sornette, D., Why Stock Markets Crash?
  Princeton University Press (2003).

\bibitem{SJ1998} Sornette, D. and A. Johansen, A hierarchical model of
  financial crashes, Physica A 261 (Nos. 3-4), 581-598 (1998).

\bibitem{JS1999} Johansen, A. and D. Sornette, Critical crashes, Risk
  12 (1), 91-94 (1999).

\bibitem{JSL1999} Johansen, A., D. Sornette and O. Ledoit, Predicting
  financial crashes using discrete scale invariance, Journal of Risk 1
  (4), 5-32 (1999).

\bibitem{JLS2000} Johansen, A., O. Ledoit and D. Sornette, Crashes as
  critical points, International Journal of Theoretical and Applied
  Finance 3 (2), 219-255 (2000).

\bibitem{SJ2001} Sornette, D. and A. Johansen, Significance of
  log-periodic precursors to financial crashes, Quantitative Finance
  1, 452-471 (2001).

\bibitem{Ide-Sornette} Ide, K. and D. Sornette, Oscillatory
  finite-time singularities in finance, population and rupture,
  Physica A 307 (1-2), 63-106 (2002).

\bibitem{Weirstrass} Zhou, W.-X. and D. Sornette, Renormalization
  group analysis of the 2000-2002 anti-bubble in the US S\&P 500
  index: Explanation of the hierarchy of five crashes and prediction,
  Physica A 330, 584-604 (2003).

\bibitem{JS_Landau2} Sornette, D. and A. Johansen, Large financial
  crashes, Physica A 245 (3-4), 411-422 (1997).

\bibitem{JS_Landau2-3} Johansen, A. and D. Sornette, Financial
  ``anti-bubbles'': Log-periodicity in Gold and Nikkei collapses,
  Int. J. Mod. Phys. C 10(4), 563-575 (1999).

\bibitem{JS_japan_eval} Johansen, A. and D. Sornette, Evaluation of
  the quantitative prediction of a trend reversal on the Japanese
  stock market in 1999, Int. J. Mod. Phys. C Vol. 11 (2), 359-364
  (2000)

\bibitem{JS_endo-exo} Johansen, A. and D. Sornette, Shocks, crashes
  and bubbles in financial markets, Brussels Economic Review (Cahiers
  economiques de Bruxelles), 49 (3/4), Special Issue on Nonlinear
  Analysis (2006) (http://ideas.repec.org/s/bxr/bxrceb.html)
  (http://arXiv.org/abs/cond-mat/0210509)

\bibitem{ZS_stat_sign-02} Zhou, W.-X. and D. Sornette, Statistical
  significance of periodicity and log-periodicity with heavy-tailed
  correlated noise, Int. J. Mod. Phys. C 13 (2), 137-170 (2002).

\bibitem{forecasting} Sornette, D. and W.-X. Zhou, Predictability of
  large future changes in major financial indices, International
  Journal of Forecasting 22, 153-168 (2006).

\bibitem{deregulation} United States Senate Permanent Subcommittee on
  Investigations, 109th Congress 2nd Session, The Role of Market
  speculation in Rising Oil and Gas Prices: A Need to Put the Cop Back
  on the Beat; Staff Report, prepared by the Permanent Subcommittee on
  Investigations of the Committee on Homeland Security and
  Governmental Affairs, United States Senate, Washington D.C., June
  27, 2006. p. 3.
  
  \bibitem{arXiv Blogger} \url{http://arxivblog.com/?p=462}
  
  \bibitem{JiangZemin} Jiang, Ze-min, Reflections on energy issues in China,
  Journal of Shanghai Jiaotong University 42 (3), 345-359 (March 2008)
  (in chinese, but english abstract and figure captions in english).
  
  \bibitem{IEA08} International Energy Agency, Oil Market Report, Issue 13 May 2008,
\url{www.oilmarketreport.org} and \url{http://omrpublic.iea.org/}


\end{thebibliography}
\end{document}